\documentclass[aps,prl,preprint,superscriptaddress]{revtex4}

\usepackage{graphicx}
\usepackage{epsfig}
\usepackage{dcolumn}
\def\scgw{{sc$GW$}}

\def\GLDA{{G^{\rm LDA}}}

\def\WLDA{{W^{\rm LDA}}}

\def\GWLDA{${G^{\rm LDA}}{W^{\rm LDA}}$}
\def\GWA{\emph{GW}A}

\def\ei{\varepsilon_i}

\def\ej{\varepsilon_j}

\def\DVo{{\it \Delta}V(\omega)}
\def\DVo{{\it \Delta}V}

\def\H0{H^0}
\def\G0{G^0}
\def\veff{V^{\rm eff}}
\def\vxc{V^{\rm xc}}
\def\vext{V^{\rm ext}}
\def\vh{V^{\rm H}}

\newcommand{\bfr}{{\bf r}}

\def\veff{V^{\rm eff}}

\def\qpscgw{{\rm QPsc}$GW$}

\begin{document} 

\title{Quasiparticle Self-Consistent {\em GW} Theory}

\author{M. van Schilfgaarde}
\affiliation{Arizona State University, Tempe, AZ, 85287}

\author{Takao Kotani}
\affiliation{Arizona State University, Tempe, AZ, 85287}

\author{S.~Faleev}
\affiliation{Sandia National Laboratories, Livermore, CA 94551}


\begin{abstract}

In past decades the scientific community has been looking for a reliable
first-principles method to predict the electronic structure of solids with
high accuracy.  Here we present an approach which we call the quasiparticle
self-consistent $GW$ approximation (\qpscgw).  It is based on a kind of
self-consistent perturbation theory, where the self-consistency is
constructed to minimize the perturbation.  We apply it to selections from
different classes of materials, including alkali metals, semiconductors,
wide band gap insulators, transition metals, transition metal oxides,
magnetic insulators, and rare earth compounds.  Apart some mild exceptions,
the properties are very well described, particularly in weakly correlated
cases.  Self-consistency dramatically improves agreement with experiment,
and is sometimes essential.  Discrepancies with experiment are systematic,
and can be explained in terms of approximations made.

\end{abstract}

\maketitle

\begin{figure}[htbp]
\centering
\epsfig{file=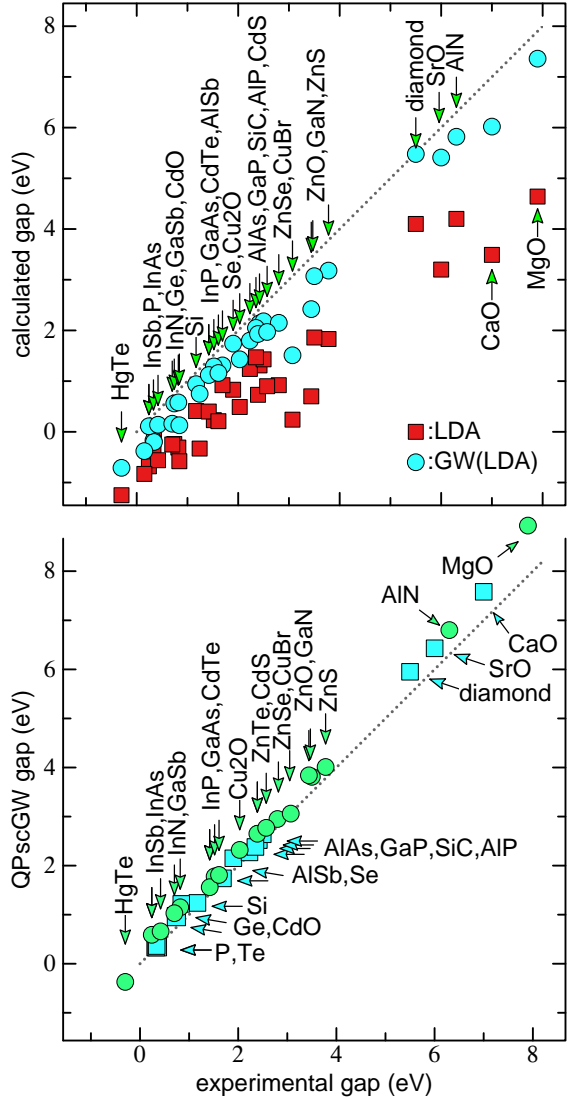,angle=0,width=0.45\textwidth,clip}
\caption{Fundamental gaps of $sp$ compounds from LDA (squares) and
  $\GLDA\WLDA$ (circles) in top panel, and from
  \qpscgw, Eqn.~(\ref{eq:veff}), in bottom panel.  The spin-orbit coupling
  was subtracted by hand from the calculations.  The $\GLDA\WLDA$ gaps
  improve on the LDA, but are still systematically underestimated.  For
  \qpscgw\ data, zincblende compounds with direct $\Gamma-\Gamma$
  transitions are shown as green circles; All other gaps are shown as blue
  squares.  Errors are small and highly systematic, and would be smaller
  than the figure shows if the electron-phonon renormalization were
  included,}
\label{fig:gaps}
\end{figure}

\begin{figure}[htbp]
\epsfig{file=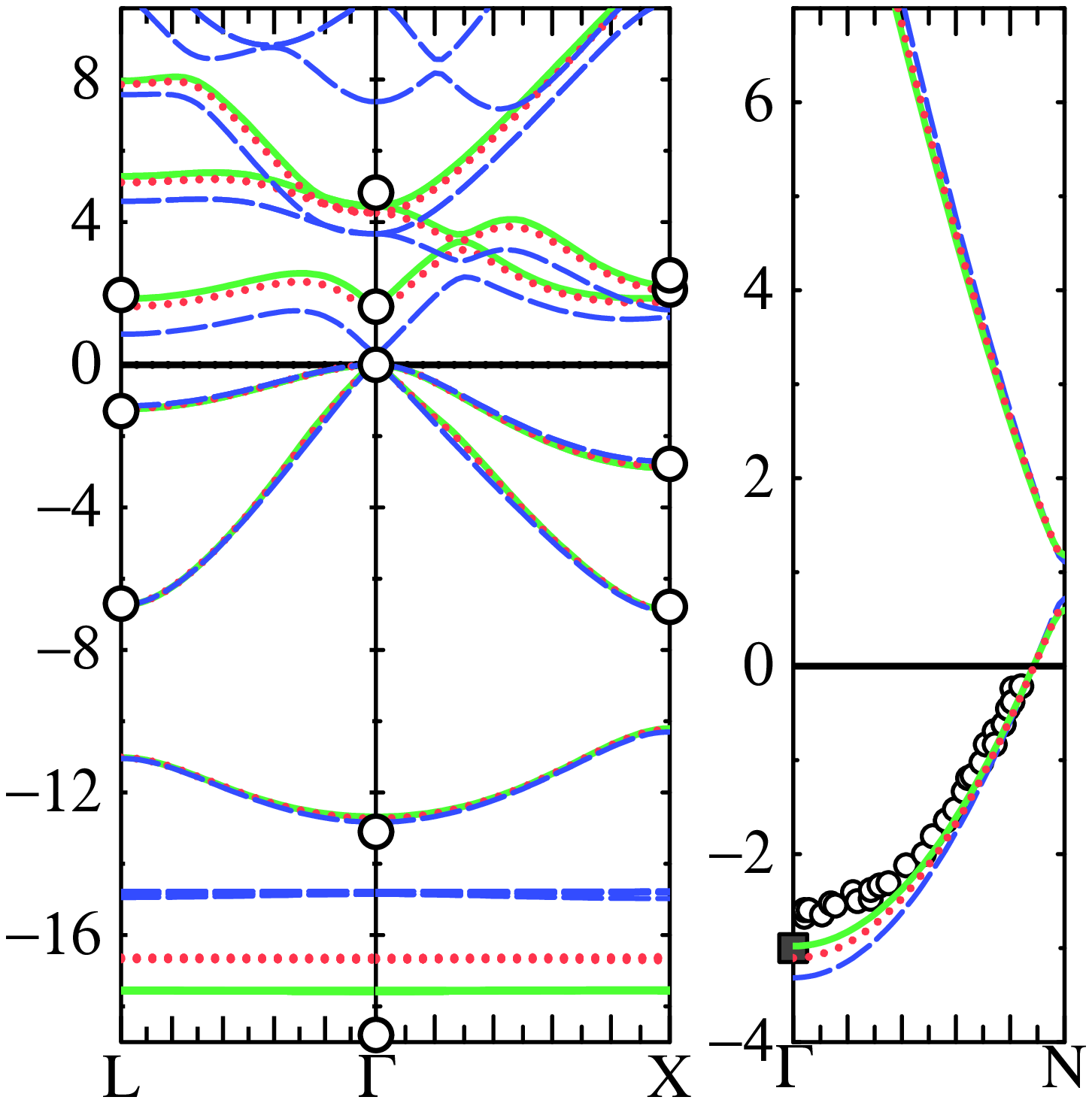,angle=0,width=0.45\textwidth,clip}
%
\caption{Comparison of LDA (blue dashes), $\GLDA\WLDA$ (red
  dots) and {\qpscgw} (green lines) energy bands in GaAs (left) and Na (right).
  Circles are experimental data, with spin-orbit coupling
  subtracted by hand.
  The \qpscgw\ fundamental gap and conduction-band effective mass 
  ($E_g=1.77\,$eV and $m_c^{*}=0.077\,m_0$) are slightly overestimated, and the optical
  dielectric constant underestimated ($\epsilon_\infty=8.4$):
  $E_g^{\rm expt}(0{\rm K})=1.52\,$eV, $m_c^{*,\rm{}expt}=0.065\,m_0$, and
  $\epsilon^{\rm expt}_\infty=10.8$.
  For comparison, $E_g^{\rm \GLDA\WLDA}$=1.29~eV and $m_c^{*,\rm
  \GLDA\WLDA}\approx{}0.059m_0$, while $E_g^{\rm{}LDA}=0.21\,$eV and
  $m_c^{*,\rm LDA}=0.020\,m_0$.
  The correspondence between {\qpscgw} and  experiment at other known
  levels at $\Gamma$, L, and X, the Ga $3d$ level near
  $-18\,$eV is representative of nearly all available data for $sp$
  systems.  For Na, the 
 {\qpscgw} occupied bandwidth is 15\% smaller than the LDA.  Circles taken from
  photoemission data~\cite{lyo88}; square from momentum electron
  spectroscopy~\cite{vos02}.}
\label{fig:gasbands}
\end{figure}

The Schr\"odinger equation is the fundamental equation of condensed matter,
and the importance of being able to solve it reliably can hardly be
overestimated.  The most widely used theory in solids, and now in quantum
chemistry, is the celebrated local density approximation
(LDA)\cite{kohn65}.  In spite of its successes, it is well known that the
LDA suffers from many deficiencies, even in weakly correlated materials
(see Figs.~\ref{fig:gaps} and \ref{fig:gasbands}).  This has stimulated the
development of flavors of extensions to the LDA to redress one or another
of its failures, such as the LDA+$U$ method.  Each of these methods
improves one failing or another in the LDA, but they often bear a
semi-empirical character, and none can be considered universal and
parameter-free.  Thus we are far from a precise and universally applicable
theory for solids, with attendant limits their ability to predict materials properties.

The random phase approximation (RPA) or {\em{}GW} approximation ($GW$A,
{\em{}G}=Green's function, {\em{}W}=screened coulomb interaction) of
Hedin\cite{hedin65} is almost as old as the LDA.  A major advance was put
forward by Hybertsen and Louie\cite{hybertsen86} when they employed LDA
eigenfunctions to generate the {\em{}GW} self-energy $\Sigma=iGW$, and
showed that fundamental gaps in $sp^{3}$ bonded materials were considerably
improved over the LDA.  Since that seminal work, many papers and some
reviews\cite{ferdi98,aulbur00,onida02} have been published on $GW$ theory
and extensions to it.  One problem that has plagued the {\em{}GW} community
has been that calculated results of the same quantities tend to vary
between different groups, much as what occurred in the early days of the
LDA.  This is because \emph{further} approximations are usually employed
which significantly affect results.  Almost ubiquitous is the 1-shot
approximation where (following Hybertsen) $\Sigma\approx{}i\GLDA\WLDA$:
{\em i.e.}, $\Sigma$ is computed from LDA eigenfunctions.  However, there
is an emerging
consensus\cite{kotani02,weiku02,alouani03,tiago04,Fleszar05,gwadequacy}
that, when cores are treated adequately\cite{gwadequacy}, $\GLDA\WLDA$
bandgaps are underestimated even in (weakly correlated) semiconductors.
(see top panel of Fig.~\ref{fig:gaps}; note especially CuBr).



In general, one-shot $GW$ approaches are rather unsatisfactory.  The
quality of the $\GLDA\WLDA$ approximation is closely tied to the quality of
LDA starting point\cite{gwadequacy}, and is adequate to construct $G$ and
$W$ only under limited circumstances\cite{gwadequacy}.  It can fail even
qualitatively in transition-metal and rare earth compounds such as CoO and
ErAs\cite{gwadequacy}.  Some kind of self-consistency is essential: the QP
levels should not be an artifact of the starting conditions.  The full
self-consistent $GW$ method (full {\scgw}) determines $G$ self-consistently
from $\Sigma=iGW$, which in turn generates $G$.  Here $W=v(1-vP)^{-1}$
where $P=\!-iG\times{}G$ and $v$ are respectively the irreducible
polarization function and (bare) Coulomb interaction.  In the few cases
where it has been applied, some difficulties were found: in particular the
valence bandwidth of the homogeneous electron gas\cite{holm98} is
$\sim$15\% {\em wider} than the noninteracting case, whereas the $\GLDA
\WLDA$ width is $\sim$15\% {\em narrower}, in agreement with experiment for
Na (see Fig.~\ref{fig:gasbands}).  A recent (nearly) full {\scgw} study of
Ge and Si also overestimates the valence bandwidth\cite{weiku02}, though
the fundamental gaps are well described.


In Ref.~\cite{faleev04} we proposed an ansatz for a different kind of
{\scgw}, and demonstrated that it radically improves the quasiparticle (QP)
levels in the oxides MnO and NiO.  In this Letter, we
ground the idea on an underlying principle---namely optimization of
the effective one-body hamiltonian $\H0$ by minimizing the perturbation to
it---and propose it as a universal approach to the reliable prediction of
the electronic structure.  We show that this approach, which we call the
quasiparticle self-consistent $GW$ (\qpscgw) method, results in accurate
predictions of excited-state properties for a large number of weakly and
moderately correlated materials.  QP levels are uniformly good for all
materials studied: not just fundamental gaps in semiconductors but for
nearly all levels where reliable experimental data are available.  Even in
strongly correlated $d$ and $f$ electron systems we studied, errors are
somewhat larger but still systematic.

The {\GWA} is usually formulated as a perturbation theory starting from a
non-interacting Green's function $\G0$ for given one-body hamiltonian
$\H0=\frac{-\nabla^2}{2m}+\veff$. $\H0$ is noninteracting, so $\veff$ is
static and hermitian but it can be nonlocal.  Because the {\GWA} is an
approximation to the exact theory, the one-body
effective hamiltonian
$H(\omega)=\frac{-\nabla^2}{2m}+\vext+\vh+\Sigma(\omega)$ depends on
$\veff$ and is a functional of it: the Hartree potential $\vh$ is generated
through $\G0={1}/\left({\omega-\H0 \pm i \epsilon}\right)$, and the {\GWA}
generates $\Sigma(\omega)$.  $H(\omega)$ determines the
time-evolution of the one-body amplitude for the many-body system.

{\qpscgw} is a prescription to determine the optimum $\H0$: we choose
$\veff$ based on a self-consistent perturbation theory so that the
time-evolution determined by $\H0$ is as close as possible to that
determined by $H(\omega)$, within the RPA.  This idea means that we have to
introduce a norm $M$ to measure the difference
$\Delta{}V(\omega)=H(\omega)-\H0$; the optimum $\veff$ is then that
potential which minimizes $M$.  A physically sensible choice of norm is
\begin{eqnarray}
M[\veff] &=& {\rm Tr}\left[\DVo \delta(\omega - \H0) \{\DVo\}^{\dagger}\right] \nonumber \\
         &+& {\rm Tr}\left[\{\DVo\}^{\dagger} \delta(\omega - \H0)\DVo\right]\ 
\label{eq:norm}
\end{eqnarray}
where the trace is taken over $\bfr$ and $\omega$.  Exact minimization $M$
is apparently not tractable, but an approximate solution can be found.
Note that $M$ is positive definite.  If we neglect the second term and
ignore the restriction that $\veff$ is hermitian, we have the trivial minimum
$M[\veff]=0$ at $\veff = V^{\rm ext}+V^{\rm H}+\vxc$ where $\vxc =
\sum_{ij} |\psi_i\rangle \Sigma(\ej)_{ij} \langle\psi_j|$.  Here
$\Sigma(\ei)_{ij}= \langle\psi_i|\Sigma(\ei)|\psi_j\rangle$, and
$\{\psi_i,\epsilon_i\}$ are eigenfunctions and eigenvalues of $\H0$.  The
second term is similarly minimum with $\Sigma(\ei)\to\Sigma(\ej)$.  An
average of the hermitian parts of these solutions results in
\begin{eqnarray}
\vxc = \frac{1}{2}\sum_{ij} |\psi_i\rangle 
       \left\{ {{\rm Re}[\Sigma(\ei)]_{ij}+{\rm Re}[\Sigma(\ej)]_{ij}} \right\}
       \langle\psi_j|.
\label{eq:veff}
\end{eqnarray}
{\rm Re} signifies the hermitian part.  This result is
the same as Eq.~(2) in Ref.~\onlinecite{faleev04}.  

We identify solutions to $\H0$ as ``bare QP'', which interact via the
(bare) $v$.  The dressed QP consists of the central bare QP plus induced
polarized clouds of the other bare QPs'---this is nothing but the physical
picture in RPA to calculate poles of $G$ from $\G0$.  In the charged Fermi
liquid theory \cite{pines66} of Landau and Silin, the QP interact via $v$
in addition to the short-range Landau interaction ($f_{pp'}$ in
Ref.~\onlinecite{pines66}; see Eq.~(3.41)).  We can virtually construct the
Landau-Silin QP from $\G0$ by the calculation of the non-RPA contributions
to $\Sigma$.  Or we can identify our bare QP as the Landau-Silin QP if we
assume they are minimally affected by such contributions.

Our {\qpscgw} is conceptually very different from the full {\scgw}.  In the
latter case the electron-hole mediated state making $P=-iG\times{}G$ is
suppressed by the square of the renormalization factor $Z \times Z$, and
includes physically unclear contributions such as: QP $\times$ (incoherent
parts)~\cite{faleev04,Bechstedt97}.  $P$ loses its physical meaning as the
density response function, $P=\delta n/\delta V$, but is merely an
intermediate construction in the self-consistency cycle.  Such a
construction does not give reasonable $W$ even in the electron gas
\cite{holm98,tamme99}, resulting in a poor $G$.


We now turn to {\qpscgw} results, focusing on the QP energies given by
$\H0$.  Fig~\ref{fig:gasbands} shows that the \qpscgw\ valence band in Na
properly \emph{narrows} relative to the LDA by 15\%.  Indeed, for nearly
all the $sp$ semiconductors studied, calculated QP levels generally agree
very closely with available experimental data.  The best known are the
fundamental gaps, shown in Fig.~\ref{fig:gaps}.  {\qpscgw} data is divided
into circles for materials whose gap is a $\Gamma-\Gamma$ transition and
squares for all other kinds.  Roughly, $\Gamma-\Gamma$ transitions are
overestimated by 0.2$\,$eV, while the remaining gaps are overestimated by
0.1$\,$eV.  Errors appear to be larger for wide-gap, light-mass compounds
(bearing elements C, N, and especially O); however, the calculations omit
reduction in the gaps by the nuclear zero-point motion.  This effect has
been studied through varying isotopic mass in some tetrahedral
semiconductors\cite{cardona04}.  It is largest for light compounds: $T$=0
the gap is reduced by $\sim$0.3 eV in diamond and $\sim$0.2~eV in AlN, but
$\lesssim$0.1~eV for heavier compounds.  Because the renormalization been
measured only for a few cases, we do not include it here.

%


Apart from some mild exceptions, \qpscgw\ generates a consistently precise
description of the electronic structure in $sp$ systems, including other
known excitations.  This is illustrated in Fig.~\ref{fig:gasbands}, where
GaAs was chosen because of the abundance of available experimental data.
It is notable that the errors are not only small, but unlike the {\GWLDA}
or LDA, they are highly systematic: compare, for example, the fundamental
gaps (Fig.~\ref{fig:gaps}).  We may expect that the bandgaps should be
overestimated, because the RPA dielectric function omits electron-hole
correlation effects.  Thus $\epsilon^{RPA}$ should be too small and
under-screen $W$.  Indeed, the optical dielectric constant
$\epsilon_\infty$ is systematically underestimated slightly
(Fig.~\ref{fig:gasbands}).  Similar consistencies are found in the
effective masses.  The conduction-band mass at $\Gamma$,
$m^{*,{\rm{}QPsc}GW}_{c\Gamma}$ consistently falls within a few percent of
experimental data for wide-gap materials, but as the gap becomes smaller
(induced by, e.g. scaling $\Sigma$),
$m^{*,{\rm{}QPsc}GW}_{c\Gamma}/m^{*,{\rm{}expt}}_{c\Gamma}$ scales
essentially as the ratio of the {\qpscgw} gap to the experimental one, as
expected when the gap become small.  Taking data for GaAs from
Fig.~\ref{fig:gasbands} for example, we obtain $E_g^{{\rm QPsc}GW}/E_g^{\rm
expt}=1.16$, and
$m^{*,{\rm{}QPsc}GW}_{c\Gamma}/m^{*,{\rm{}expt}}_{c\Gamma}=1.18$.



\begin{table}[htbp] \caption{ \label{tab:threed} Valence $d$
bandwidths $W_d$ (calculated at $\Gamma$ for Ti,Cr, and Co, and at N for
Fe, and at X for Ni), relative position of $s$ and $d$ band bottoms
$\epsilon_{sd}$, splittings $\Delta{}E_{\rm x}$ between majority
and minority $d$ (or $f$) states, and magnetic moments in $3d$
compounds and Gd. 
}


\centering

  \begin{tabular}{l lll|lll}
                & \multicolumn{3}{c}{$W_d$~(eV)}      &    \multicolumn{3}{c}{$\epsilon_{sd}$~(eV)}  \\
                & LDA     & \qpscgw     & Expt & LDA     & \qpscgw     & Expt \\
     Ti         & 6.0     &  5.7        &      & 3.5     &  4.3        &      \\
     Cr         & 6.6     &  6.2        &      & 3.5     &  4.3        &      \\
     Fe         & 5.2     &  4.6        & 4.6  & 3.6     &  4.4        & 4.6  \\
     Co         & 4.1     &  3.8        & 3.7  & 4.6     &  5.3        & 4.9\rlap{$\pm1$}  \\
     Ni         & 4.4     &  4.0        & 4.0  & 4.4     &  5.0        & 5.5  \\
\hline
                & \multicolumn{3}{c}{moment ($\mu_B$)}     &    \multicolumn{3}{c}{$\Delta{}E_{\rm x}$~(eV)}  \\
                & LDA     & \qpscgw     & Expt                 & LDA     & \qpscgw    & Expt \\
     Fe         & 2.2     &  2.2        & 2.2              & 1.95        &  1.67      & 1.75 \\
     Co         & 1.6     &  1.7        & 1.6              & 1.70        &  1.21      & 1.08 \\
     Ni         & 0.6     &  0.7        & 0.6              & 0.6         &  0.5       & 0.3\hbox{ } \\
     MnO        & 4.5     &  4.8        & 4.6  \\
     NiO        & 1.3     &  1.7        & 1.9  \\
     MnAs       & 3.0     &  3.5        & 3.4  \\
     Gd         & 7.7     &  7.8        & 7.6              & 4.9         &  16.1      & \hskip -8 pt{$\sim$}12.1
  \end{tabular}
  \end{table}

Table~\ref{tab:threed} shows that the $3d$ bandwidth, the relative position
of $s$ band, exchange splittings $\Delta{}E_{\rm x}$, are systematically improved
relative to the LDA in elemental $3d$ metals, and Gd.  \qpscgw\ magnetic
moments are systematically overestimated slightly.  $\Delta{}E_{\rm x}$ is overestimated in
Ni, presumably owing to the neglect of spin fluctuations\cite{ferdi92}.  
\qpscgw\ also
predicts with reasonable accuracy the QP levels of all magnetic $3d$ compounds
studied, in particular correlated oxides such as MnO and NiO
where the LDA fails dramatically.  As might be expected, the accuracy
deteriorates somewhat relative to $sp$ systems.  For example, the \qpscgw\
optical gap in NiO ($4.8\,$eV) was found to be larger than experiment
($\sim$4.3$\,$eV).  Table~\ref{tab:threed} compares the magnetic moments,
and Ref.~\cite{faleev04} shows in detail the QP levels are consistently
well described.  However, for Gd, (and for GdP and GdAs)
{\qpscgw} overestimates the position of the (empty) minority Gd $f$
shell by $\sim$4~eV, and hence the exchange splitting $\Delta{}E_{\rm x}$.

\begin{figure}[htbp]
\centering
\epsfig{file=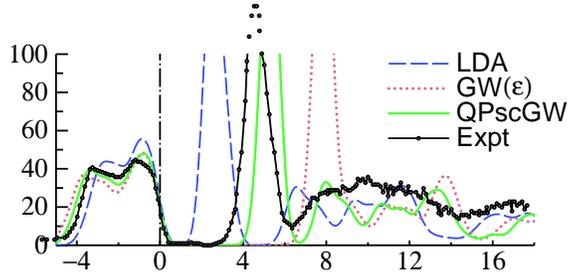,angle=0,width=0.45\textwidth,clip}
\caption{DOS in CeO$_2$.  Black dots are PES+BIS
data[\cite{wuilloud84}].  Calculated DOS were broadened with
Gaussian of width 0.35~eV.  $\GLDA\WLDA$+eigenvalue-only self-consistency
(dotted red line) severely overestimates the position of the Ce $f$ level,
while it is slightly overestimated by \qpscgw (green line).  Broadening
of the valence bands relative to LDA is found in all oxides studied,
e.g. MgO and TiO$_2$, and has important consequences, e.g. in determining valence-band offsets.}
\label{fig:ceo2}
\end{figure}

Nonmagnetic oxides SrTiO$_3$, TiO$_2$, and CeO$_2$, with conduction bands
of $d$ or $f$ character, overestimate fundamental gaps slightly more than
their $sp$ counterparts.  The \qpscgw\ gaps were found to be $4.19\,$eV and
$3.78\,$eV in SrTiO$_3$ and TiO$_2$, $\sim$0.8$\,$eV larger than the
experimental gaps ($\sim$3.3$\,$eV and $\sim$3.1$\,$eV).
Fig.~\ref{fig:ceo2} compares the \qpscgw\ DOS of CeO$_2$ with spectroscopic
data: the Ce $f$ band is similarly overestimated by \qpscgw.  This is
reasonable, because electron-hole correlation effects are stronger in the
narrow $d$ ($f$) conduction bands.  Fig.~\ref{fig:ceo2} also shows DOS
computed by $\GLDA\WLDA$, but with eigenvalue-only self-consistency, where
only the diagonal part in Eq.~(\ref{eq:veff}) is kept.  This constrains the
eigenfunctions to the starting (LDA) eigenfunctions; thus the charge and
spin densities do not change.  While the off-diagonal parts of $\Sigma$ add
a small effect in, e.g. GaAs, their contribution is essential in CeO$_2$,
even though the occupied states contain only a small amount of Ce $f$
character.

To summarize, the {\qpscgw} theory (apart from some mild exceptions)
appears to be an excellent predictor of QP levels for a variety materials
selected from the entire periodic table.  Self-consistency is an essential
part of the theory.  From the results obtained so far, this approach shows
promise to be universally applicable scheme, sufficient in its own right
for QP levels in many materials.  In contrast to the LDA or any other
popular theory of electronic structure of solids in the literature today,
the method is truly \emph{ab initio} with errors that are generally small
and highly systematic across many different materials classes.  The errors
can be attributed missing electron-hole correlation contributions to
$\epsilon$.  When better calculations are necessary (usually where the
physics lies completely outside the domain of a one-particle picture, such
as the description of excitons, multiplets, or Mott transitions), \qpscgw\
can be taken as an optimum starting point where the relevant many-body
contributions to the hamiltonian are (nearly) as small as possible.  The
systematic character of the error suggests that the dominant terms left out
can be described by a few diagrams, in particular the ladder diagrams
coupling electrons and holes; the smallness of the error suggests that the
additional terms can be added as a perturbation around the \qpscgw\ $\H0$,
without the need for further self-consistency.


This work was supported by ONR contract N00014-02-1-1025 and BES
Contract No. DE-AC04-94AL85000.






\end{document}